\documentclass[fleqn,usenatbib]{mnras}

\usepackage{mathptmx}

\usepackage[T1]{fontenc}

\DeclareRobustCommand{\VAN}[3]{#2}
\let\VANthebibliography\thebibliography
\def\thebibliography{\DeclareRobustCommand{\VAN}[3]{##3}\VANthebibliography}

\usepackage{graphicx}	
\usepackage{amsmath}	
\usepackage{amssymb}	
\usepackage{subfigure}  
\usepackage{caption}    
\usepackage{hyperref}   

\title[Evolution of phase space densities in SFR]{The evolution of phase space densities in star-forming regions}

\author[G. A. Blaylock-Squibbs \& R. J. Parker]{
George A. Blaylock-Squibbs\thanks{E-mail: gablaylock-squibbs1@sheffield.ac.uk} and Richard J. Parker\thanks{Royal Society Dorothy Hodgkin fellow}
\\
Department of Physics and Astronomy,
The University of Sheffield, Hounsfield Road,
Sheffield,
S3 7RH\\
}

\date{Accepted XXX. Received YYY; in original form ZZZ}

\pubyear{2022}

\begin{document}
\label{firstpage}
\pagerange{\pageref{firstpage}--\pageref{lastpage}}
\maketitle

\begin{abstract}
The multi-dimensional phase space density (both position and velocity) of star-forming regions may encode information on the initial conditions of star and planet formation. Recently, a new metric based on the Mahalanobis distance has been used to show that hot Jupiters are more likely to be found around exoplanet host-stars in high \textit{6D} phase space density, suggesting a more dynamic formation environment for these planets. However, later work showed that this initial result may be due to a bias in the age of hot Jupiters and the kinematics of their host stars. We test the ability of the Mahalanobis distance and density to differentiate more generally between star-forming regions with different morphologies by applying it to static regions that are either substructured or smooth and centrally concentrated. We find that the Mahalanobis distance is unable to distinguish between different morphologies, and that the initial conditions of the N-body simulations cannot be constrained using only the Mahalanobis
distance or density. Furthermore, we find that the more dimensions in the phase space the less effective the Mahalanobis density is at distinguishing between different initial conditions. We show that a combination of the mean three-dimensional (x, y, z) Mahalanobis density and the $Q$-parameter for a region can constrain its initial virial state. However this is due to the discriminatory power of the $Q$-parameter and not from any extra information imprinted in the Mahalanobis density. We therefore recommend continued use of multiple diagnostics for determining the initial conditions of star-forming regions, rather than relying on a single multi-dimensional metric.
\end{abstract}




\section{Introduction}
\label{sec:introduction}
Star formation is observed to take place along filaments within giant molecular clouds \citep[][]{palmeirim_herschel_2013, schisano_identification_2014, andre_interstellar_2017}. The initial formation and distribution of these filaments is likely due to supersonic turbulence within GMCs \citep{larson_turbulence_1981}. It is along these filaments that cores can form, with further fragmentation of these cores leading to stars forming in groups containing  10s to 1000s of members \citep[][]{lada_embedded_2003, bastian_spatial_2009}. 

One of the foundational questions of star formation is whether star formation is a universal process or not. Are the initial conditions of star-forming regions dependant on the environment, where differences in the stellar density, IMF and stellar multiplicity are due to the initial conditions of the star-forming region? Or does star formation happen in a similar way everywhere, and any differences we observe in these regions is stochastic in nature?

There are two main proposed modes of star-formation, monolithic and hierarchical. In monolithic formation modes the gas is already contained within the final volume of the region before stars begin to form, whereas in hierarchical formation the gas extends beyond the final volume of the region \citep[][]{longmore_formation_2014, williams_initial_2022}. In the hierarchical mode stars are forming while at the same time the gas is collapsing. 

The kinds of star-forming regions these modes produce is of interest not only in the context of star formation but also the way in which the final star-forming regions that form may influence the architecture of the planetary systems that are produced within them \citep[][]{adams_birth_2010_test, parker_birth_2020}.

Star-formation is a rapid process, occurring within a few crossing times \citep[][]{elmegreen_star_2000}, which is often less than 1 Myr. During this process, stars are forming and moving \citep[][]{alcock_mass_2019}, further muddying the formation picture. And whilst observations of the earliest stages have improved greatly with e.g. ALMA, observations of star-forming regions are often at older ages, where significant dynamical evolution may have taken place \citep[][]{klessen_mean_2001, allison_early_2010, parker_dynamical_2014, daffern-powell_dynamical_2020, schoettler_dynamical_2019}. Dynamical evolution alters the spatial and kinematic distributions of young stars, erasing the signature of the initial conditions, but can be used as a proxy for age and used to converge on a likely set of initial conditions for a given star-forming region \citep[][]{parker_dynamical_2014}. To enable comparisons of observations and simulations, we need to be able to quantify parameters of the star-formation regions, such as the degree of substructure and mass segregation \citep[][]{cartwright_statistical_2004, allison_using_2009, sanchez_spatial_2009, alfaro_looking_2016, gonzalez_phase-space_2017, jaffa_mathcal_2017, buckner_spatial_2019, arnold_quantifying_2022, joncour_multiplicity_2018, kuhn_spatial_2014, gouliermis_complex_2014}. 


Early methods such as the auto-correlation function and two-point correlation function compared the number of excess star pairings to a random distribution of stars as a function of scale \citep[][]{gomez_spatial_1993, larson_star_1995}. These methods where used extensively to determine the degree of substructure, with early work suggesting breaks in the two-point correlation function corresponded to the Jeans length \citep[][]{simon_clustering_1997} (though see \citet[][]{bate_interpreting_1998}) and the size of the widest stellar binaries in the regions in question \citep[][]{kraus_spatial_2008, joncour_multiplicity_2017}. 

Subsequent work made extensive use of minimum spanning trees (MSTs) to quantify structures in star-forming regions. \citet[][]{cartwright_statistical_2004} introduced the $Q$-parameter to quantify spatial substructure, and \citet[][]{allison_using_2009} introduced the $\Lambda_{\rm MSR}$ method to quantify mass segregation.

\citet[][]{parker_dynamical_2014} showed that the initial conditions of a region can be inferred from the spatial information, if a suitable number of metrics are combined, including the relative stellar surface density around the most massive stars \citep[][]{maschberger_global_2011, kupper_mass_2011}.

However, the majority of the above methods are designed to operate on two or three-dimensional spatial data, whereas recent observational data (e.g. from Gaia and associated ground-based surveys) has provided high resolution spatial and kinematic data (\textit{6D}).

Recently, in an attempt to quantify the phase space densities of exoplanet host stars \citet[][]{winter_stellar_2020} developed the Mahalanobis density, a new application of the Mahalanobis distance \citep[][]{mahalanobis_generalized_1936}. 

The Mahalanobis distance has been used in astronomy for classifying objects, for example in \citet[][]{siegal_multivariate_1974} it is used to analyse and classify different types of asteroid impact craters and in \citet[][]{jakimiec_multivariate_1991} it was used to classify sunspots into groups. 

Due to the differing dimensions, and units of very different scale (i.e. length in pc and velocity in kms$^{-1}$) making multivariate comparisons can be difficult. However the Mahalanobis distance makes multivariate comparisons possible over wide ranges of dynamical scales by rescaling the axes and removing the 
units. \citet[][]{winter_stellar_2020} used this method to develop the Mahalanobis density and use it to propose the hypothesis that host stars in high phase space densities are more likely to have hot Jupiter planets ($M > 50 \,\, M_{\rm \bigoplus}$ and $a < 0.2$ au) around them compared to the lower phase space densities. However, \citet[][]{mustill_hot_2022} show that this result may be due to a bias from the peculiar velocities of the stars. When the peculiar velocities of the stars are accounted for, there is no longer an excess of hot Jupiters in high \textit{6D} (x, y, z, Vx, Vy, Vz) phase space densities.


Irrespective of the ongoing debate surrounding the application of the Mahalanobis distance to exoplanet host stars, in this paper we aim to test this metric when applied to both synthetic static regions and assess its performance in quantifying phase space structures of N-body simulations of star-forming regions.

The paper is structured as follows. In \S~\ref{sec:methods} we present the methods used. In \S~\ref{subsec:static region} we present the results of testing the Mahalanobis distance's ability to differentiate between different morphologies. In \S~\ref{subsec:nbody results} we show how the Mahalanobis distance and density change with time in N-body simulations. In \S~\ref{subsec:comparing to established methods} we compare the  Mahalanobis distance and density to other methods for quantifying structure in star-forming regions. In \S~\ref{sec:discussion} we present a discussion of our results and we conclude in \S~\ref{sec:conclusion}.

\section{Methods}
\label{sec:methods}
In this section we describe the set-up of the N-body simulations, including the initial spatial and kinematic distributions of stars. We then describe the methods used to quantify the spatial and kinematic distributions in our simulated star-forming regions.

\subsection{$N$-Body Simulations}
\label{sec:n-body simulations}

\begin{figure}
    \centering
    \subfigure[1 pc radius $D_{\rm f}=1.6$]{\includegraphics[width=0.49\linewidth]{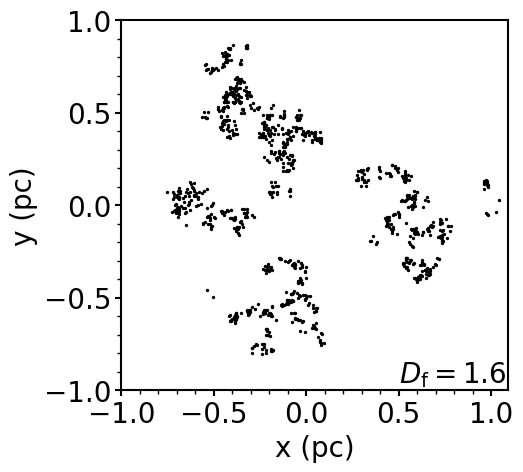}}
    \hspace{0.1pt}
     \subfigure[1 pc radius $D_{\rm f}=3.0$]{\includegraphics[width=0.49\linewidth]{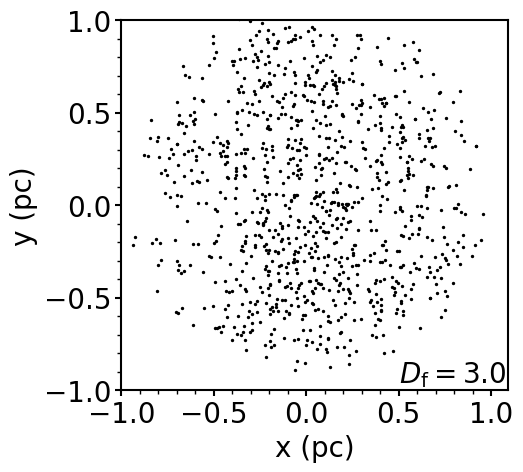}}
    \caption{Examples of fractal regions with 1000 stars, on the left-hand side is a highly substructure region of fractal dimension $D_{\rm f}=1.6$ and on the right-hand side is a region with far less substructure with fractal dimension $D_{\rm f}=3.0$.}
    \label{fig:frac example region}
\end{figure}

The N-body simulations are run using the \texttt{Kira} integrator, part of the \texttt{Starlab}\footnote{\href{https://www.sns.ias.edu/~starlab/index.html}{https://www.sns.ias.edu/\textasciitilde starlab/index.html}} package \citep[][]{zwart_star_1999, portegies_zwart_star_2001}. Each simulation has a population of 1000 stars and is run for a simulated time of 10 Myr. Our choice of 1000 stars comes from \citet[][]{lada_embedded_2003} where they find the following relation,
\begin{equation}
    N_{\rm cl} \propto M_{\rm cl}^{-2}
\end{equation}
where $N_{\rm cl}$ is the number of clusters and $M_{\rm cl}$ is the mass of the cluster. This power-law is obeyed for star clusters between masses of $10 < M_{\rm cl}/M_{\rm \odot} < 10^{5}$ and our choice of 1000 stars puts our simulations close to the middle of this distribution.



Observations of star forming regions show that within these complexes there are filaments of denser gas, inside of which prestellar cores are observed \citep[][]{myers_filamentary_2009, andre_interstellar_2017}. These filaments are thought to be caused by supersonic turbulence, which is likely to be responsible for the substructure we observe in these regions \citep[][]{larson_turbulence_1981, kraljic_role_2014}. To mimic this substructure we make use of a box-fractal generator to initialise our simulations with fractal dimensions of $D_{\rm f}=1.6$ and $D_{\rm f}=3.0$ \citep[][]{goodwin_dynamical_2004, daffern-powell_dynamical_2020}.

We run subvirial simulations to match observations that indicate prestellar cores may be subvirial with respect to one another and may then virialise and form bound smooth, centrally concentrated star clusters \citep[][]{foster_-sync_2015, kuznetsova_et_al_2015, parker_mass_2016}. We also run a set of simulations that are initially supervirial. We do this to mimic the observations that some young star-forming regions ($\sim 1-5$ Myr) are supervirial and therefore expanding \citep{bravi_et_al_2018, kuhn_kinematics_2019, kounkel_dynamical_2022}.

Figure~\ref{fig:frac example region} shows example clusters with the fractal dimensions $D_{\rm f}=1.6$ and $D_{\rm f}=3.0$.

We use the following to define the virial ratio, 
\begin{equation}
    \alpha_{\rm vir} = \frac{T}{|\Omega|},
\end{equation}
where $T$ is the total kinetic energy of the region and $\Omega$ is the total potential energy of the region. By using this equation we can scale the initial velocities to our desired virial ratio either $\alpha_{\rm vir} = 0.1$ for subvirial regions or $\alpha_{\rm vir} = 0.9$ for supervirial regions. The initial conditions for the simulations are summarised in table~\ref{tab:simulations run list inital conditions}.

We assign masses using the Maschberger IMF with a lower mass limit of $0.01 \,\, M_{\rm \odot}$, upper mass limit of $50.0 \,\, {M_{\rm \odot}}$ and a mean mass of $0.2 \,\, M_{\rm \odot}$ \citep[][]{maschberger_function_2013}. 

The probability distribution function of the Maschberger IMF is,
\begin{equation}
    p(m) \propto \left(\frac{m}{\mu} \right)^{-\alpha} \left(1 + \left(\frac{m}{\mu}\right)^{1 - \alpha}\right)^{-\beta},
    \label{eq:maschberger_imf}
\end{equation} 
where $\mu$ is the mean stellar mass, $\alpha = 2.3$ is the high mass exponent and $\beta = 1.4$ is the low mass exponent \citep[][]{salpeter_luminosity_1955}. 

\begin{table}
    \centering
    \caption{This table shows the different initial conditions of the simulations. For each of these initial conditions 10 simulations are run for 10 Myr. From left to right the columns are, the initial fractal dimension of the region, the number of stars, the initial virial ratio and the  initial radius of the simulations in pc.}
    
    \begin{tabular}{l|cccc}
    \hline
    
        Fractal Dimension & $N_{\bigstar}$ & Virial Ratio & Radius (pc) \\ \hline
        
        $D_{\rm f}=1.6$ & 1000 & 0.1, 0.9 & 1, 5 \\

        $D_{\rm f}=3.0$ & 1000 & 0.1, 0.9 & 1, 5 \\

        \hline
    \end{tabular}
    \label{tab:simulations run list inital conditions}
\end{table}

\subsection{Smooth Centrally Concentrated Regions}
\label{subsec:radial regions}
We generate smooth, centrally concentrated regions with radial density profiles in which the stars are randomly distributed using the following relation,
\begin{equation}
\label{eq:radial density profile}
    n \propto r^{\rm -\alpha},    
\end{equation}
where $r$ is the distance from the centre of the region and $\alpha$ is the radial density index and has values $\alpha = 0.0, 1.0, 2.0$ and $2.5$.
We also generate Plummer spheres, regions with a three-dimensional density distribution of the form,
\begin{equation}
    \rho_{\rm p}(r) = \frac{3M_0}{4\pi a^3} \left(1+\frac{r^2}{a^2}\right)^{-\frac{5}{2}},
    \label{plummer density profile}
\end{equation}
where $M_0$ is the total mass of the region, $r$ is the distance from the centre of the region and $a$ is the Plummer radius \citep[]{plummer_problem_1911, kroupa_dense_2008}.

\subsection{Generating Fractal Regions}
\label{sec:making fractal regions}
We follow \citet[][]{goodwin_dynamical_2004} and \citet[][]{cartwright_statistical_2004} to generate substructured regions using the box-fractal method. Other examples of this method can be found in (e.g. \citet[][]{allison_using_2009}, \citet[][]{parker_comparisons_2015}, \citet[][]{daffern-powell_dynamical_2020}.
 
The method proceeds as follows. A single star is placed at the centre of a cube whose side length is chosen to be $N_{\rm Div}$. This cube is then subdivided down into $N^3_{\rm Div}$ sub-cubes. A star is then placed at the centre of each of the sub-cubes. Each of these sub-cubes then has a probability of being subdivided again given by $N^{(3-D_{\rm f})}_{\rm Div}$, where $D_{\rm f}$ is the fractal dimension. Cubes that are not subdivided have their stars removed along with any previous generations of stars that came before them. A small amount of noise is added to each of the stars to prevent them having a regular looking appearance. 

These steps are repeated until the desired number of stars is reached or exceeded in the latest generation. Once this condition is met all previous generations of stars are removed, then the remaining stars are randomly removed until the desired number of stars is reached. By removing the stars in this manner we end up with stars distributed inside of a spherical volume. 

The velocities of the first generation of stars are picked from a Gaussian with mean zero, with each subsequent generation of stars inheriting the velocity of the previous generation plus a random component. This results in stars that are close to each other having similar velocities and stars far apart from one another having very different velocities. The velocities are related to the length scale of the region with the following relation $V(L) \propto L^{3-D_{\rm f}}$ where $L$ is the length scale of interest and $D_{\rm f}$ is the desired fractal dimension \citep{parker_evolution_2018}.

\subsection{The Mahalanobis Distance}
\label{sec:mahalanobis density}
The Mahalanobis distance is a metric that measures distances between points to the average in the distribution in $N$ dimensional phase spaces \citep[][]{mahalanobis_generalized_1936}. 



The Mahalanobis distance does this by removing any correlations in the data by multiplying the distances between points and the average of the region by the inverse of the covariance matrix; this also has the effect of re-scaling the data.

Once this re-scaling has been done the Euclidean distances are found in the phase space; this is the Mahalanobis distance, $M_{\rm d}$. 

Each point in a dataset is described using a vector where each element is a measured parameter of that point,
\begin{equation}
    \Vec x = \left(x_1, x_2, x_3, ... \, , x_{\rm N} \right) ^ {\rm T},
\end{equation}
where $x_1, x_2, x_3, ..., x_n$ are the parameters. For example, if each point has the three parameters, (x, y, z) then this is simply its physical position in \textit{3D} space.

The Mahalanobis distance between a point in a distribution and the mean of that distribution in an $N$ dimensional phase space is defined as,
\begin{equation}
    M_{\rm d}(\Vec{x}, \Vec{\mu}) = \sqrt{\left(\Vec{x}-\Vec{\mu}\right)^{\rm{T}} \textbf{S}^{-1} \left(\Vec{x}-\Vec{\mu}\right)},
\end{equation}
where the $\Vec{x}$ is the point vector, $\Vec{\mu}$ is a vector of the averages of the parameters of interest and $\textbf{S}^{-1}$ is the inverse of the covariance matrix for all the parameters in the region.

The Mahalanobis distance, $M_{\rm d}$ has been also used to define a parameter space density called the Mahalanobis density, $\rho_{\rm m,N}$ \citep[][]{winter_stellar_2020}\footnote{We have used different notation for the Mahalanobis distance to avoid confusion with the fractal dimension and also the number of dimensions. We instead use $M_{\rm d}$ instead of $D$ as used in \citet[][]{winter_stellar_2020} to avoid confusion with the fractal dimension. We also change the number of dimensions in the phase space from $D$ to $D_{\rm p}$ again to avoid confusion with the fractal dimension, which we represent as $D_{\rm f}$.}.

To calculate the Mahalanobis density we first must define the Mahalanobis distance between points in the phase space (i.e. distance between $\Vec{x}$ and $\vec{y}$). We follow \citet[][]{winter_stellar_2020} and use,
\begin{equation}
    m_{\rm d}(\Vec{x}, \Vec{y}) = \sqrt{\left(\Vec{x}-\Vec{y}\right)^{\rm{T}} \textbf{S}^{-1} \left(\Vec{x}-\Vec{y}\right)},
\end{equation}
where $\Vec{x}$ is the vector describing the measurements of one point, $\Vec{y}$ is the vector describing the measurements of another and $\textbf{S}^{-1}$ is the inverse of the covariance matrix of all the parameters of interest.

The calculation of the Mahalanobis density proceeds as follows. First we find the Mahalanobis distance to the $N^{th}$ nearest neighbour, then we divide the nearest neighbour number by the volume whose side length is defined as the Mahalanobis distance to the $N^{th}$ nearest neighbour. The Mahalanobis density is then defined as,
\begin{equation}
\label{eq:mahalanobis density from winter et al 2020}
    \rho_{\rm{m,N}} = N m_{\rm d, N}^{-D_{\rm p}},
\end{equation}
where $\rho_{\rm{m,N}}$ is the Mahalanobis density, $N$ is the nearest neighbour number, $m_{\rm d,N}$ is the Mahalanobis distance to the $N^{th}$ nearest neighbour and $D_{\rm p}$ is the number of dimensions in the phase space \citep{winter_stellar_2020}. The Mahalanobis densities are then normalised so that the median Mahalanobis density is unity. 

In this work we apply the Mahalanobis density to two different phase spaces, the positional phase space (\textit{3D}) and the position-velocity phase space (\textit{6D}). For this work we find the Mahalanobis distance to the 20$^{th}$ nearest neighbour in the phase space (i.e. $N = 20$), the same as in \citet[][]{winter_stellar_2020}.

\subsection{Local Surface density ratio}
\label{sec:sigma ldr}
The local surface density ratio $\Sigma_{\rm LDR}$ was introduced in \citet[][]{maschberger_global_2011} to quantify the differences between the surface densities of subsets of stars within their host regions and for this work we choose the 10 most massive stars as the subset of interest. The algorithm proceeds as follows. For each star we find the distance to its $N^{\rm th}$ nearest neighbour, then we calculate the circular area whose radius is the distance to the $N^{\rm th}$ nearest neighbour. To find the surface density of the stars we divide the nearest neighbour number by this area. We use a nearest neighbour number of 5 for this work.

The ratio is defined as,
\begin{equation}
    \label{eq:sigma ldr}
    \Sigma_{\rm LDR} = \frac{\Tilde{\Sigma}_{\rm subset}}{\Tilde{\Sigma}_{\rm all}}
\end{equation}
where $\Tilde{\Sigma}_{\rm subset}$ is the median surface density found for the 10 most massive stars and $\Tilde{\Sigma}_{\rm all}$ is the median surface density found for the entire region. Therefore, if $\Sigma_{\rm LDR} > 1$ the 10 most massive stars are found in areas of higher than average stellar surface density, and conversely, if $\Sigma_{\rm LDR} < 1$ then they are located in areas of lower than average surface density. The significance of any difference is quantified using a two-sample Kolmogorov-Smirnov test. Where if $p << 0.01$ we reject the null hypothesis that the 10 most massive stars share the same underlying distribution of surface densities compared to the entire region.

\subsection{Mass Segregation Ratio}
\label{sec:lambda msr}
The mass segregation ratio $\Lambda_{\rm MSR}$ was first introduced in \citet[][]{allison_using_2009} to quantify the degree of mass segregation in a star-forming region. The definition of mass segregation in this case is that the most massive stars are closer to each other than expected from the average separation of all of the stars in the region. The method makes use of minimum spanning trees (MSTs) which are graphs of points connected to each other in such a way that the total length of the tree is minimised and that all points are connected to at least one other point with no closed loops. 

This method generates a minimum spanning tree for the chosen subset of stars, for this work we use the 10 most massive stars. It will then pick 10 random stars from the region and make an MST for these random stars. We do this 200 times to calculate the mean edge length of the randomly chosen trees. The ratio is calculated using the following equation, 
\begin{equation}
    \label{eg:lambda msr}
    \Lambda_{\rm MSR} = {\frac{\left<l_{\rm average}\right>}{l_{\rm 10}}}^{+ \sigma_{5/6}/l_{\rm 10}}_{-\sigma_{1/6}/l_{\rm 10}},
\end{equation}
where $\left<l_{\rm average}\right>$ is the average edge length found for all the randomly constructed MSTs and $l_{\rm 10}$ is the edge length of the subset's MST. It is important to note that the random MSTs we construct can also contain members of the chosen subset.

If the ratio is $> 1$ then the region's 10 most massive stars are mass segregated, if the ratio is $\sim 1$ then the most massive stars are not mass segregated and if the ratio is less than 1 they are inversely mass segregated (the most massive stars are further apart than the average stars in the region). In this work we mark the value 1 to show the boundary between mass segregation and inverse mass segregation. However we follow \citet[][]{parker_comparisons_2015} and only take ratio values above 2 to be signs of mass segregation, to avoid false positives.

We follow \citet[][]{parker_spatial_2018} and calculate the uncertainty using the randomly constructed MSTs. First we order the lengths of the random MSTs and find the values that lie $1/6$ and $5/6$ of the way through this list. This gives us values which correspond to a 66 per cent deviation from the median MST length found.

\subsection{$Q$-Parameter}
\label{sec:q-parameter}
The $Q$-Parameter was introduced in \citet{cartwright_statistical_2004} to quantify and distinguish between different cluster morphologies. 

The $Q$-parameter also makes use of MSTs and proceeds as follows. First the normalised correlation length is found. This is the mean separation between all stars in a region which is then divided by the region's radius to normalise it. 

The mean edge length of the region is found by constructing an MST for the region and then finding the mean edge length. The mean edge length is normalised by diving it by $\frac{N_{\rm total} A}{N_{\rm total}-1}$, where $N_{\rm total}$ is the number of stars in the region and $A$ is the area of the region. We use the circular area (see \citet[][]{schmeja_evolving_2006, parker_spatial_2018} for a discussion on normalisation techniques), with the radius defined as the distance from the centre of mass to the most distant star. The $Q$-parameter is then defined as,
\begin{equation}
\label{eq:q-parameter}
Q = \frac{\Bar{m}}{\Bar{s}},
\end{equation}
where $\bar{m}$ is the normalised mean edge length of the MST and $\bar{s}$ is the normalised correlation length between stars.
Regions with substructured morphologies have $Q<0.8$ whereas regions with smooth, centrally concentrated morphologies have $Q>0.8$.

\section{Results}
\label{sec:results}
We show the results of the Mahalanobis distance applied to both static and N-body simulations of star-forming regions with various initial conditions. We present the \textit{3D} and \textit{6D} Mahalanobis densities calculated in the N-body simulations and compare the evolution of the Mahalanobis density over time to the other methods for quantifying spatial and kinematic distributions in star-forming regions.

\subsection{Static Regions}
\label{subsec:static region}
We first find the Mahalanobis distances between stars and the average point in the region, $\vec{\mu}$, and calculate the average Mahalanobis distance in their respective regions ($\bar{M}_{\rm d}$) for sets of synthetic and static star clusters, with each set having a different structural parameter. Each region in the set consists of 1000 stars. We calculate $\bar{M}_{\rm d}$ for substructured regions with fractal dimensions $D_{\rm f}=1.6, 2.0, 2.6, 3.0$ and clusters with radial density profile indexes, $\alpha = 0.0, 1.0, 2.0, 2.5$. We also show the results of a set of 100 Plummer spheres which have a radial density profile described by equation~\ref{plummer density profile} with a radial density index of $2.5$ (see \S~\ref{subsec:radial regions}).

\begin{figure}
    \centering
    \includegraphics[width=\columnwidth]{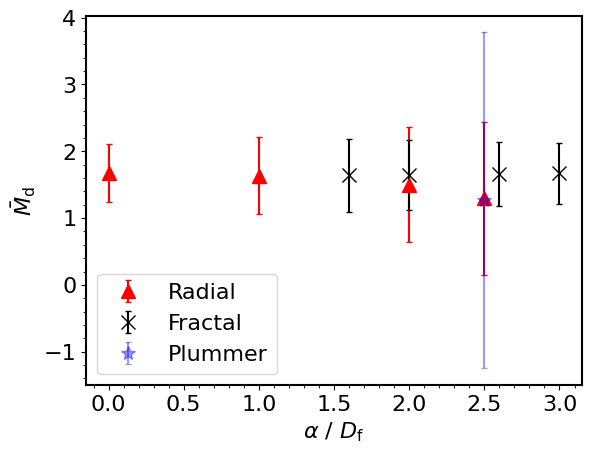}
    \caption{Mean of the mean Mahalanobis distances calculated in the \textit{3D} phase space for sets of 100 different star-forming regions plotted against the structural parameter used to make the sets. The red triangles are the smooth, centrally concentrated radial regions, the purple star (on top of the red triangle with the structural parameter equal to 2.5) is the Plummer sphere and the black crosses are the fractal regions. The error bars show a single standard deviation.}
    \label{fig:Mahalanobis distance structure plot}
\end{figure}

Figure~\ref{fig:Mahalanobis distance structure plot} shows the mean of the means for the $M_{\rm d}$ (the Mahalanobis distance of each star to its region's averages in the \textit{3D} phase space) for the 100 clusters in each set of initial conditions. We first calculate the mean Mahalanobis distance in each of the 100 regions in the set and then we calculate the mean of those means. The error bars show the standard deviation of the mean of the mean Mahalanobis distances found in each of the regions in a particular set. Figure~\ref{fig:Mahalanobis distance structure plot} clearly shows that $\bar{M}_{\rm d}$ is degenerate across a wide range of morphologies and we are therefore unable to use $\bar{M}_{\rm d}$ calculated in the \textit{3D} phase space to differentiate between the different morphologies. There is much more spread in the values for the Plummer sphere compared to the radial and fractal models, with the fractals having the smallest spread of $\bar{M}_{\rm d}$ and the radial regions sitting between the two. This is due to Plummer spheres being formally infinite in extent, so the calculation occasionally has to normalise over very distant stars.

\subsection{N-body Results}
\label{subsec:nbody results}
\begin{figure*}
    \hspace{0.8pt}
    \subfigure[$D_{\rm f}=1.6$, subvirial, 1 pc]{\includegraphics[width=0.49\linewidth]{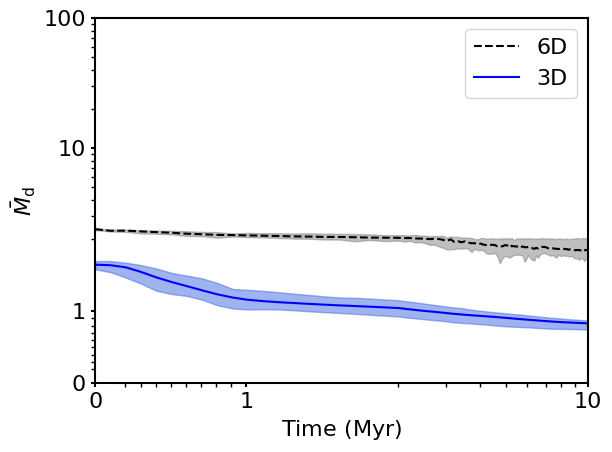}}
    \hspace{0.8pt}
    \subfigure[$D_{\rm f}=1.6$, supervirial, 1 pc]{\includegraphics[width=0.49\linewidth]{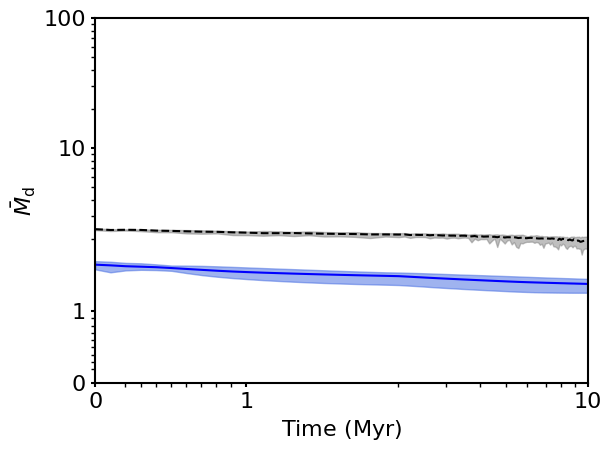}}

    
    \caption{Plots of the mean Mahalanobis distance calculated in both the \textit{3D} and \textit{6D} phase spaces against time for regions with both high initial volume densities and high degrees of substructure (i.e. fractal dimension $D_{\rm f} = 1.6$ with radii of 1 pc) consisting of 1000 stars. The shaded areas show the range of mean Mahalanobis distances found across all 10 of the simulations at the current time. The solid lines show the mean of the mean Mahalanobis distances across all 10 simulations. The blue area and solid blue line shows the \textit{3D} phase space and the black dashed line and the grey area show the same but for the \textit{6D} phase space.}
    
    \label{fig:mahadis vs time frac 1d6 1 pc}
\end{figure*}

Figure~\ref{fig:mahadis vs time frac 1d6 1 pc} shows the mean of the mean Mahalanobis distances (calculated in both \textit{3D} and \textit{6D}) found for 10 different N-body simulations with initial fractal dimension $D_{\rm f}=1.6$ and 1 pc radii. The left hand panel shows the results for subvirial (collapsing) regions, and the right hand panel shows the results for the supervirial (expanding) regions. In both the sub- and supervirial cases there is a decrease in the Mahalanobis distance over time for both the \textit{3D} and \textit{6D} phase spaces. For the initially subvirial simulations the \textit{3D} Mahalanobis distance swiftly decreases at the start and then continues to decrease for the rest of the simulation but at a slower rate. For the supervirial regions we see a less pronounced decrease in $\bar{M}_{\rm d}$ compared to the initially subvirial regions. The $\bar{M}_{\rm d}$ calculated in the \textit{6D} phase space shows more modest decrease over time for both initially sub- and supervirial simulations.

Figure~\ref{fig:mean maha v time d=1.6, 1pc} shows the mean Mahalanobis density ($\bar{\rho}_{\rm m, 20}$) calculated in both the \textit{3D} and \textit{6D} phase spaces for two sets of 10 (one subvirial and the other supervirial) simulations with an initial fractal dimension of $D_{\rm f}=1.6$ and initial radii of 1 pc. The highest Mahalanobis densities are calculated in the \textit{3D} phase space (x, y, z) for the supervirial simulations, however these large final values are only present for a few of the simulations. In the \textit{3D} phase space the Mahalanobis density increases in the first 2-4 Myr, after which the density stays the same for the rest of the run time. This is most likely due to the early dynamical interactions of stars; as they move closer to each other the stars' Mahalanobis densities will increase. In \textit{Appendix}~\ref{app: maha distance versus maha density} we show the relationship between the Mahalanobis distance and density for the high density simulations with and without substructure. We show this for both the \textit{3D} and \textit{6D} phase spaces.

Initially, in the subvirial simulations, the \textit{6D} Mahalanobis density decreases for the first 1 Myr and then stays the same until around 5 Myr where it then starts to increase again. For the supervirial simulations this initial decrease in the \textit{6D} phase space happens more rapidly than in the subvirial simulations. It also does not reach the low densities that the subvirial regions attain. In the supervirial regions, the stars will expand together in co-moving groups, therefore they will have similar positions but can still have velocities that exhibit kinematic substructure. In contrast, in the subvirial simulations the stars interact more and erase this substructure. The difference in the velocities explains why the \textit{6D} Mahalanobis density is much lower than the \textit{3D} density. Some of the supervirial regions attain higher \textit{6D} Mahalanobis densities compared to the subvirial regions at the end of the 10 Myr. 

\begin{figure*}
\hspace{0.8pt}
\subfigure[$D_{\rm f}=1.6$, subvirial, 1 pc]{\includegraphics[width=0.49\linewidth]{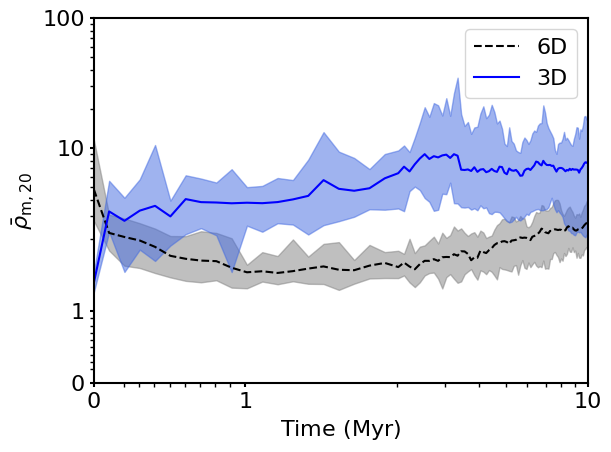}} 
\hspace{0.8pt} 
\subfigure[$D_{\rm f}=1.6$, supervirial, 1 pc]{\includegraphics[width=0.49\linewidth]{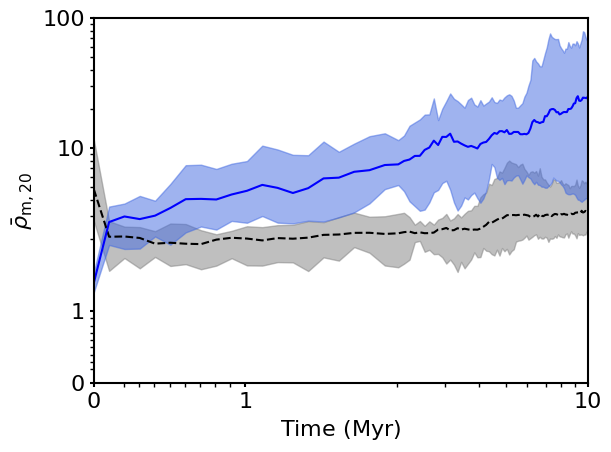}}

\caption{The mean Mahalanobis density against time for each of the 10 subvirial (left-hand panel) and supervirial simulations (right-hand panel) with fractal dimension $D = 1.6$ and radii of 1 pc. The simulations consist of 1000 stars. The blue shaded area shows the minimum and maximum mean Mahalanobis density (in the \textit{3D} phase space) found across all 10 of the simulations at the current time. The solid blue line shows the mean of the means for the Mahalanobis density in the \textit{3D} phase space. The grey shaded area and the dashed black line shows the same but for the Mahalanobis density calculated in the \textit{6D} phase space. The mean number densities of the 10 simulations is around 314 stars pc$^{-1}$ with a mean stellar mass density of around $201 \,\, M_{\rm \odot}$ pc$^{-3}$.}

\label{fig:mean maha v time d=1.6, 1pc}
\end{figure*}

We also calculate the Mahalanobis density and distance for low density simulations for simulations with $D_{\rm f} = 1.6$ and $D_{\rm f} = 3.0$, where the initial radii regions are 5 pc (with a mean number density of around 3 stars pc$^{-3}$ and a mean stellar mass density of around $1.6 \,\, M_{\rm \odot}$ pc$^{-3}$). We find that the evolution of $\bar{M}_{\rm d}$ is almost identical to the its evolution in the high density simulations, both in \textit{3D} and \textit{6D} phase spaces.

We show the evolution of the \textit{3D} and \textit{6D} Mahalanobis densities for these more diffuse simulations in Figure~\ref{fig:mean maha v time d=1.6, 5pc}, which shows $\bar{\rho}_{\rm m, 20}$ plotted against time for the same substructured regions with fractal dimension $D_{\rm f}=1.6$ with initial radii of 5 pc. For the \textit{3D} phase space we see the same trends as in Figure~\ref{fig:mean maha v time d=1.6, 1pc} but a slight difference for the \textit{6D} phase space. Now $\bar{\rho}_{\rm m, 20}$ decreases in the subvirial simulations as the regions evolve, and the supervirial simulations show a steady Mahalanobis density after 1 Myr. 

We show the evolution of the Mahalanobis distance (measured between the points and the mean values of the phase in each snapshot), and Mahalanobis density in the simulations that have no primordial substructure (i.e. they are uniform spheres at $t = 0$ Myr). 

\begin{figure*}
    \hspace{0.8pt}
    \subfigure[$D_{\rm f}=3.0$, subvirial, 1 pc]{\includegraphics[width=0.49\linewidth]{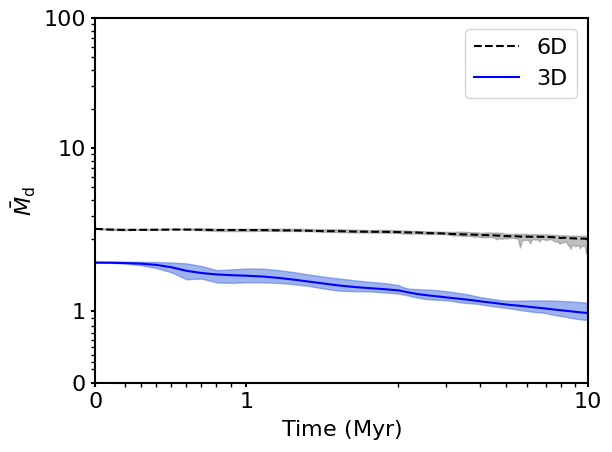}}
    \hspace{0.8pt}
    \subfigure[$D_{\rm f}=3.0$, supervirial, 1 pc]{\includegraphics[width=0.49\linewidth]{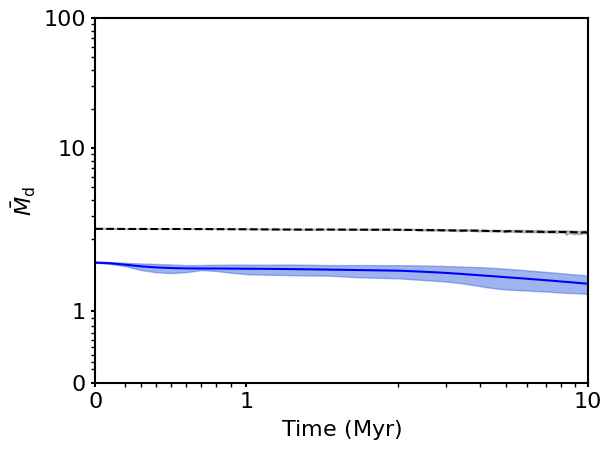}}
    
    \caption{Plots of the mean Mahalanobis distance from each star to the average in simulations without primordial substructure, i.e. a fractal dimension of $D_{\rm f} = 3.0$, over time. The shaded blue area and solid blue line show the minimum and maximum mean Mahalanobis distance found across the 10 simulations and the the solid blue line shows the mean of the means across all 10 simulations, respectively. The Mahalanobis distance is calculated in the \textit{3D} phase space for the blue area and line and calculated in the \textit{6D} phase space, shown by the grey shaded area and the black dashed line.}

    \label{fig:mahadis vs time D=3.0 1 pc}
\end{figure*}

Figure~\ref{fig:mahadis vs time D=3.0 1 pc} shows the Mahalanobis distance over time for regions with a fractal dimension $D_{\rm f}=3.0$ with radii of 1 pc. For the \textit{3D} phase space there is a decrease over time for the sub- and supervirial simulations. 

Comparing these results to Fig.~\ref{fig:mahadis vs time frac 1d6 1 pc} we see that the $D_{\rm f} = 3.0$ regions' Mahalanobis distances decrease at a slower rate compared to regions with initially more substructure. This behaviour also results in a slightly greater \textit{3D} Mahalanobis distance being measured for regions with fractal dimension $D_{\rm f}=3.0$ at 10 Myr. However, this is not seen in the \textit{6D} phase space Mahalanobis distances which show little change over the 10 Myr in the simulations. Also, very little difference is seen when comparing the \textit{6D} Mahalanobis distances between the sub- and supervirial simulations.

\begin{figure*}
\hspace{0.8pt}
\subfigure[$D_{\rm f}=1.6$, subvirial, 5 pc]{\includegraphics[width=0.49\linewidth]{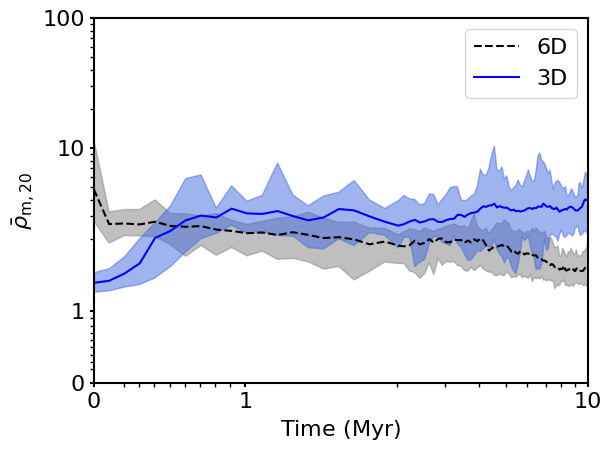}} 
\hspace{0.8pt} 
\subfigure[$D_{\rm f}=1.6$, supervirial, 5 pc]{\includegraphics[width=0.49\linewidth]{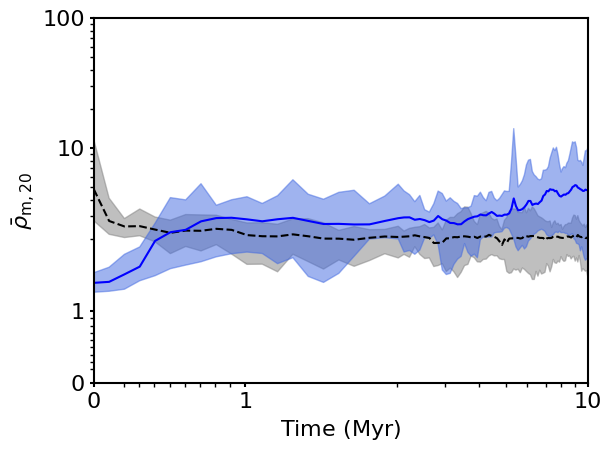}}

\caption{Plots showing the mean Mahalanobis density against time for each of the 10 subvirial (left-hand panels) and supervirial (right-hand panels) simulations with fractal dimension $D_{\rm f} = 1.6$ and radii of 5 pc. These simulations have a low initial stellar number density with a mean around 3 stars pc$^{-3}$ and a mean stellar mass density of around $1.6 \,\, M_{\rm \odot}$ pc$^{-3}$. The shaded blue area shows the minimum and maximum mean Mahalanobis density found across all 10 simulations in the \textit{3D} phase space. The solid blue line shows the mean of the means Mahalanobis density against time. The shaded grey area and the dashed black line show the same but for the \textit{6D} phase space.}
\label{fig:mean maha v time d=1.6, 5pc}
\end{figure*}

Figure~\ref{fig:mean maha v time d=3.0, 1pc} shows the \textit{3D} and \textit{6D} $\bar{\rho}_{\rm m, 20}$ against time for the simulations with no primordial substructure (with an initial fractal dimension $D_{\rm f}=3.0$ and radii 1 pc). The left-hand panel shows $\bar{\rho}_{\rm m, 20}$ against time for the subvirial simulations. It shows the same increase in Mahalanobis density as the $D_{\rm f}=1.6$ simulations but we don't see the same initial decrease in the \textit{6D} $\bar{\rho}_{\rm m, 20}$ that we see in the $D_{\rm f}=1.6$ simulations. At around 0.5 Myr a decrease in the \textit{3D} Mahalanobis density is seen, then a second period of increasing Mahalanobis density is seen around 0.9 Myr before attaining a steady Mahalanobis density for the rest of the simulations' run time. This initial increase is due to the region collapsing and stars moving closer to each other which raises the \textit{3D} Mahalanobis density. What stops it increasing further is likely the dynamical interactions causing stars to move further away from each other. Once this initial dynamical stage settles down the density can increase again due to stars being close to each other near the centre of the region. 

The right-hand panel of Figure~\ref{fig:mean maha v time d=3.0, 1pc} shows the \textit{3D} and \textit{6D} Mahalanobis density calculated for regions that are initially supervirial. The \textit{`bump'} like feature is much smaller for the \textit{3D} phase space calculations than in the initially subvirial simulations. The decrease in the bump compared to the subvirial simulations is due to the fact that the stars are constantly and continuously moving away from each other, meaning that the slight increase that is still present is due to small groupings of stars clumping together. We see similar behaviour for the \textit{6D} phase space in both the subvirial and supervirial simulations.

\begin{figure*}
\hspace{0.8pt}
\subfigure[$D_{\rm f}=3.0$, subvirial, 1 pc]{\includegraphics[width=0.49\linewidth]{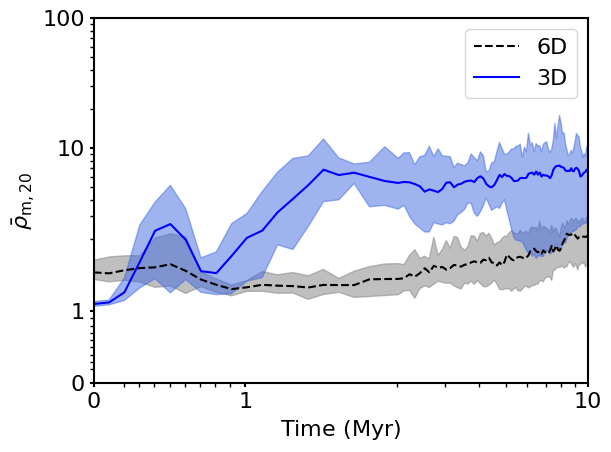}} 
\hspace{0.8pt} 
\subfigure[$D_{\rm f}=3.0$, supervirial, 1 pc]{\includegraphics[width=0.49\linewidth]{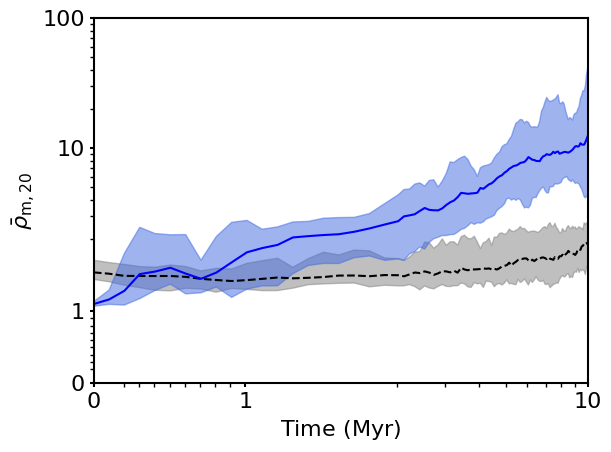}} 
\hspace{0.8pt}

\caption{The mean Mahalanobis density against time for each of the 10 subvirial (left-hand panel) and supervirial (right-hand panel) simulations without primordial substructure (fractal dimension $D_{\rm f} = 3.0$) and radii of 1 pc. The shaded blue area and solid blue line show the range of mean Mahalanobis densities calculated in the \textit{3D} phase space and the mean of the means found across all 10 simulations, respectively. The grey shaded area and the dashed black line show the same but for the \textit{6D} phase space.}
\label{fig:mean maha v time d=3.0, 1pc}
\end{figure*}

\begin{figure*}
\hspace{0.8pt}
\subfigure[$D_{\rm f}=3.0$, subvirial, 5 pc]{\includegraphics[width=0.49\linewidth]{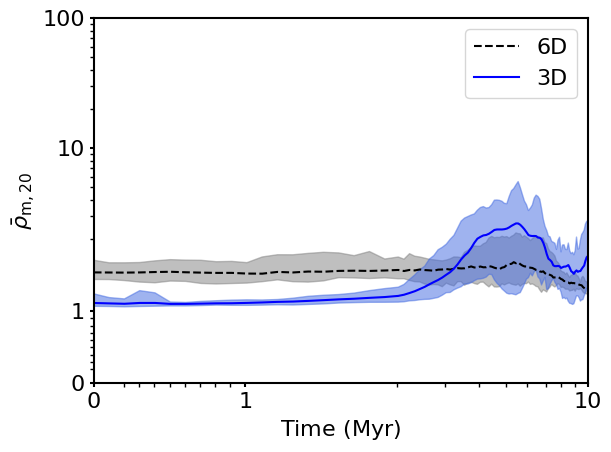}} 
\hspace{0.8pt} 
\subfigure[$D_{\rm f}=3.0$, supervirial, 5 pc]{\includegraphics[width=0.49\linewidth]{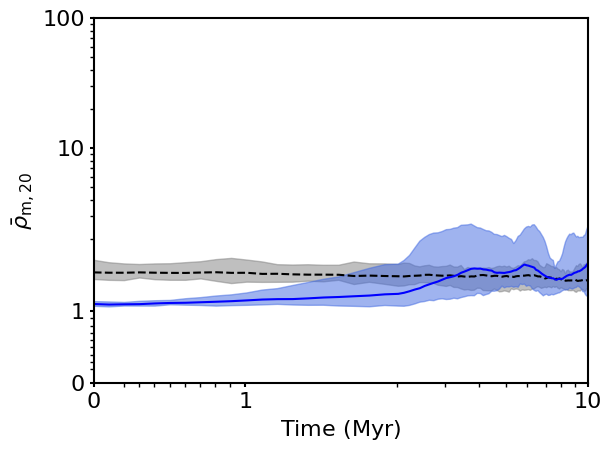}} 
\hspace{0.8pt}

\caption{The mean Mahalanobis density against time for each of the 10 subvirial (left-hand panel) and supervirial (right-hand panel) simulations without primordial substructure (fractal dimension $D_{\rm f} = 3.0$) with radii of 5 pc. The shaded blue area and solid blue line show the range of mean Mahalanobis densities calculated in the \textit{3D} phase space and the mean of the means found across all 10 simulations, respectively. The grey shaded area and the dashed black line show the same but for the \textit{6D} phase space.}
\label{fig:mean maha v time d=3.0, 5pc}
\end{figure*}

We show the \textit{3D} and \textit{6D} Mahalanobis densities for simulations with $D_{\rm f} = 3.0$ and initial radii of 5 pc in Figure~\ref{fig:mean maha v time d=3.0, 5pc}. These low density simulations have mean number density of around 1 star pc$^{-3}$, or mean stellar mass density of around $0.7 \,\, M_{\rm \odot}$ pc$^{-3}$. We find similar results to the more dense subvirial simulations simulations, where the \textit{3D} Mahalanobis density clearly traces the \textit{`bump'} of the collapse, which then decreases as stars move apart. The time the \textit{`bump'} occurs is delayed by several Myr compared to the higher density regions, due to the longer dynamical time scales.

\subsection{Comparison to other methods of quantifying structure}
\label{subsec:comparing to established methods}

\begin{figure*}
\hspace{0.8pt}
\subfigure[$\bar{\rho}_{\rm m, 20}$ vs $\Lambda_{\rm MSR}$, subvirial, 1 pc]{\includegraphics[width=0.49\linewidth]{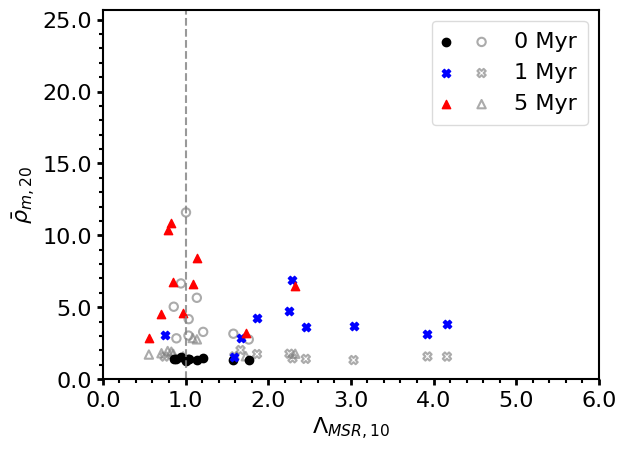}} 
\hspace{0.8pt} 
\subfigure[$\bar{\rho}_{\rm m, 20}$ vs $\Lambda_{\rm MSR}$, supervirial, 1 pc]{\includegraphics[width=0.49\linewidth]{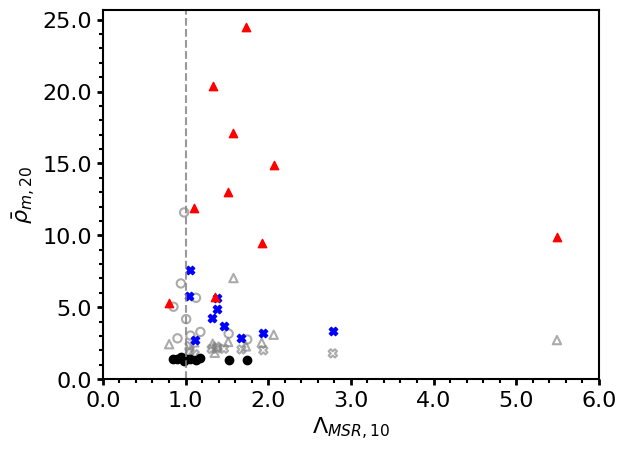}} 
\hspace{0.8pt}
\subfigure[$\bar{\rho}_{\rm m, 20}$ vs $Q$, subvirial, 1 pc]{\includegraphics[width=0.49\linewidth]{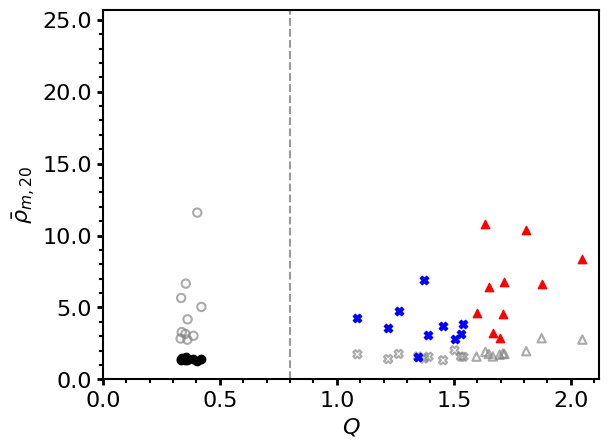}} 
\hspace{0.8pt} 
\subfigure[$\bar{\rho}_{\rm m, 20}$ vs $Q$, supervirial, 1 pc]{\includegraphics[width=0.49\linewidth]{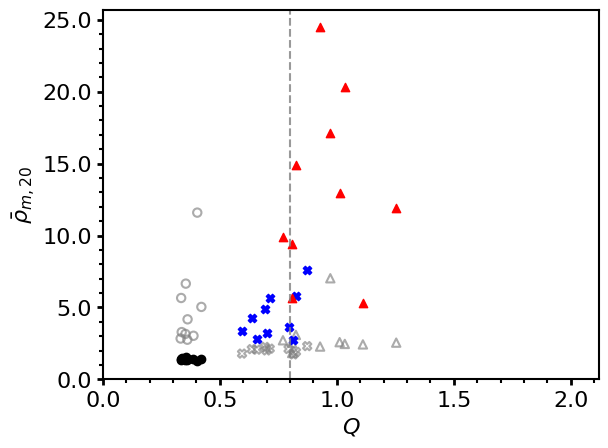}}
\hspace{0.8pt}
\subfigure[$\bar{\rho}_{\rm m, 20}$ vs $\Sigma_{\rm LDR}$, subvirial, 1 pc]{\includegraphics[width=0.49\linewidth]{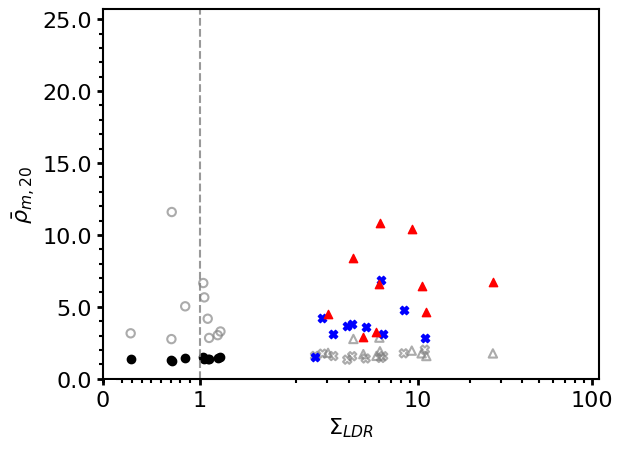}} 
\hspace{0.8pt} 
\subfigure[$\bar{\rho}_{\rm m, 20}$ vs $\Sigma_{\rm LDR}$, supervirial, 1 pc]{\includegraphics[width=0.49\linewidth]{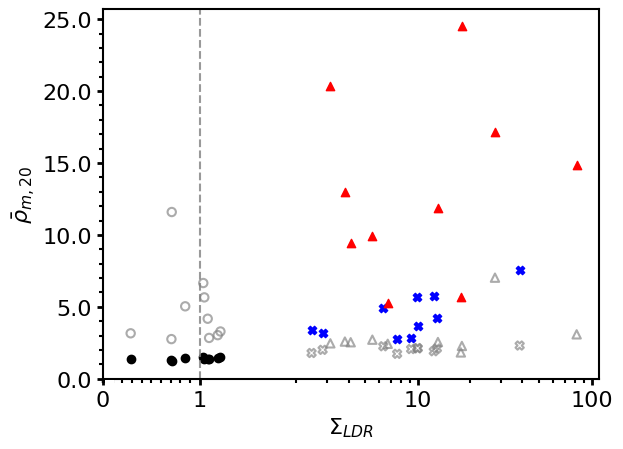}}
\hspace{0.8pt}
\caption{The mean Mahalanobis density calculated for \textit{3D} and \textit{6D} phase spaces plotted against other methods of quantifying structure for 10 subvirial and supervirial simulations which are initially substructured with fractal dimension $D_{\rm f}=1.6$ and 1 pc radii. The left-hand panels show the results for the subvirial regions and the right-hand panels show the results for the supervirial regions. The initial values at 0 Myr are represented by the black circles, the blue crosses show 1 Myr and the red triangles show 5 Myr. The grey open circles show the comparison of the \textit{6D} Mahalanobis density at 0 Myr, the open grey crosses show it for 1 Myr and the open grey triangles for 5 Myr. From top to bottom the rows show the different methods which $\bar{\rho}_{\rm m, 20}$ is plotted against, with the top row showing $\Lambda_{\rm MSR}$, second row showing $Q$ and the bottom row showing $\Sigma_{\rm LDR}$.}
\label{fig:mahaden comp plots D=1.6 1pc 3D1}
\end{figure*}

We now plot the \textit{3D} and \textit{6D} Mahalanobis densities against other measures of quantifying structure in star-forming regions.
Figure~\ref{fig:mahaden comp plots D=1.6 1pc 3D1} shows $\bar{\rho}_{\rm m, 20}$ plotted against the established methods of  $\Lambda_{\rm MSR}$, $Q$ and $\Sigma_{\rm LDR}$ for the simulations with initial fractal dimension $D_{\rm f}=1.6$ and radius 1 pc. For the initially subvirial simulations the $\bar{\rho}_{\rm m, 20}$ values stay below 12 for the first 5 Myr whereas the supervirial simulations can achieve much higher values. This is due to the stars in supervirial regions forming small groupings as the region expands which causes an increase in the Mahalanobis density. In contrast, for the subvirial regions more stars are interacting with each other, which erases spatial and kinematic substructure and also ejects stars \citep[][]{schoettler_runaway_2020}. As we are measuring the mean Mahalanobis density we are sensitive to a small number of stars being ejected which we see as a decrease in the mean Mahalanobis density for the subvirial simulations.

We first show the Mahalanobis density versus the amount of mass segregation as defined by $\Lambda_{\rm MSR}$ \citet[][]{allison_using_2009} in Figure~\ref{fig:mahaden comp plots D=1.6 1pc 3D1}\textit{(a)} and \textit{(b)}. For the subvirial simulations mass segregation is detected for 6 of the 10 simulations at 1 Myr and only one simulation has mass segregation present at 5 Myr, with $\Lambda_{\rm MSR} > 2$. The reason for the dissipation in the amount of mass segregation is due to the ejection of massive stars from unstable Trapezium-like systems \citep[][]{allison_early_2010, allison_formation_2011, parker_comparisons_2015}. In the supervirial simulations (see panel~\textit{(b)}) one region becomes mass segregated at 1 Myr and another at 5 Myr. If the cluster splits in two, with the most massive stars located in one of the halves then $\Lambda_{\rm MSR}$ can increase to the value we see in Figure~\ref{fig:mahaden comp plots D=1.6 1pc 3D1} \textit{(b)} of around 5.5. As discussed in \citet{parker_dynamical_2014}, this is because the massive stars generally do not interact with each other as they do in the subvirial simulations where there is more mixing resulting in any structure in the phase spaces being erased. 

The supervirial simulations display a wider spread in the Mahalanobis densities meaning that the plot of $\bar{\rho}_{\rm m, 20}$ versus $\Lambda_{\rm MSR}$ can be used to distinguish between different initial virial states after at least 5 Myr of dynamical evolution.

The clearest distinction between different times in the simulations comes when $\bar{\rho}_{\rm m ,20}$ is combined with the $Q$-parameter. Figure~\ref{fig:mahaden comp plots D=1.6 1pc 3D1}\textit{(c)} and \textit{(d)} show this clearly for both the subvirial simulations and the supervirial simulations. The plots also show that, as expected, the supervirial simulations maintain substructure for longer. With some of the regions maintaining traces of substructure for 5 Myr as measured using $Q$ (i.e. $Q < 0.8$). 

Panels \textit{(e)} and \textit{(f)} of Fig.~\ref{fig:mahaden comp plots D=1.6 1pc 3D1} show $\bar{\rho}_{\rm m, 20}$ plotted against the relative local surface density ratio, $\Sigma_{\rm LDR}$. We find that for both subvirial and supervirial simulations there is an increase in the local surface density of the 10 most massive stars compared to all stars in the region. Interestingly the simulations with the highest local surface density around the 10 most massive stars do not necessarily have the highest Mahalanobis densities. This is likely due to the local surface density being calculated on the plane of the sky whereas the Mahalanobis density is being calculated for the full \textit{3D} phase space.

The supervirial regions display high Mahalanobis densities at later times, and we can use this, and the different evolution of the $Q$-parameter and $\Lambda_{\rm MSR}$ to distinguish between initial conditions after several Myr of dynamical evolution.

Using the \textit{6D} Mahalanobis density does not improve the diagnostic ability of the metric. We find that the range of $\bar{\rho}_{\rm m, 20}$ decreases when calculated in \textit{6D}. Meaning we see that the $\bar{\rho}_{\rm m, 20}$ values overlap making differentiating between different snapshots and virial states impractical.

\begin{figure*}
\hspace{0.8pt}
\subfigure[$\bar{\rho}_{\rm m, 20}$ vs $\Lambda_{\rm MSR}$, subvirial, 1 pc]{\includegraphics[width=0.49\linewidth]{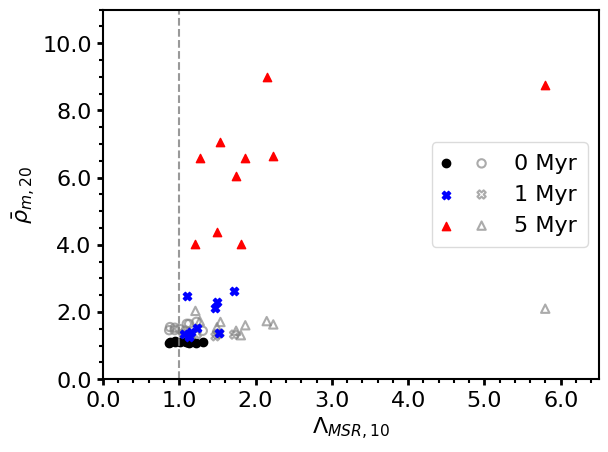}} 
\hspace{0.8pt} 
\subfigure[$\bar{\rho}_{\rm m, 20}$ vs $\Lambda_{\rm MSR}$, supervirial, 1 pc]{\includegraphics[width=0.49\linewidth]{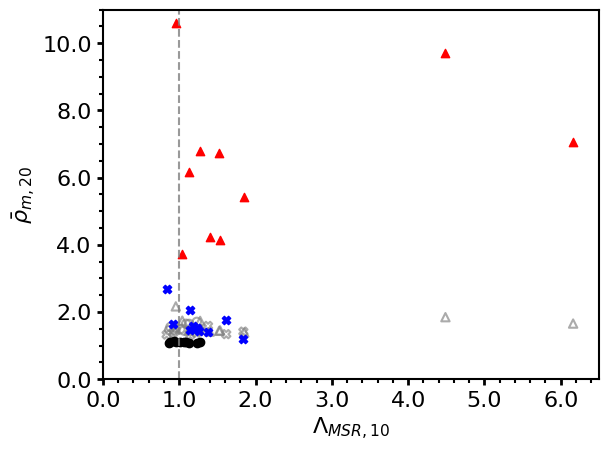}} 
\hspace{0.8pt} 
\subfigure[$\bar{\rho}_{\rm m, 20}$ vs $Q$, subvirial, 1 pc]{\includegraphics[width=0.49\linewidth]{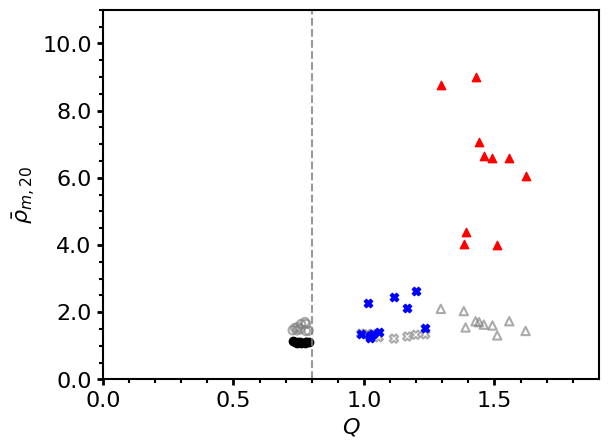}} 
\hspace{0.8pt} 
\subfigure[$\bar{\rho}_{\rm m, 20}$ vs $Q$, supervirial, 1 pc]{\includegraphics[width=0.49\linewidth]{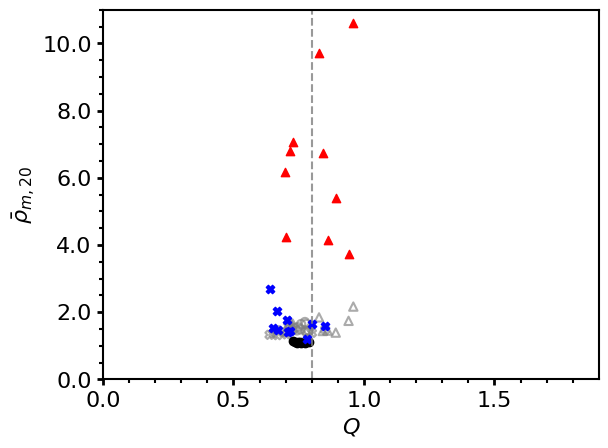}} 
\hspace{0.8pt} 
\subfigure[$\bar{\rho}_{\rm m, 20}$ vs $\Sigma_{\rm LDR}$, subvirial, 1 pc]{\includegraphics[width=0.49\linewidth]{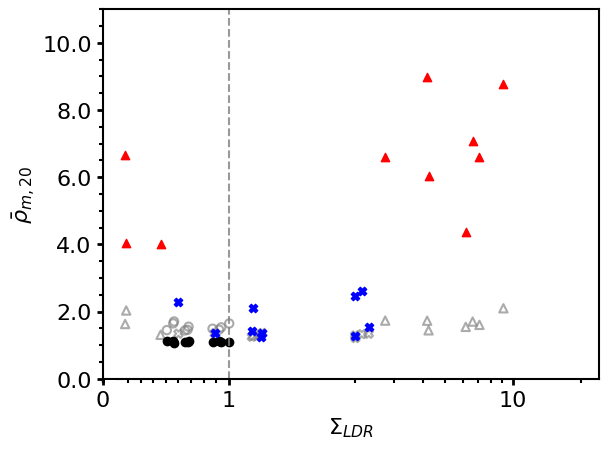}} 
\hspace{0.8pt} 
\subfigure[$\bar{\rho}_{\rm m, 20}$ vs $\Sigma_{\rm LDR}$, supervirial, 1 pc]{\includegraphics[width=0.49\linewidth]{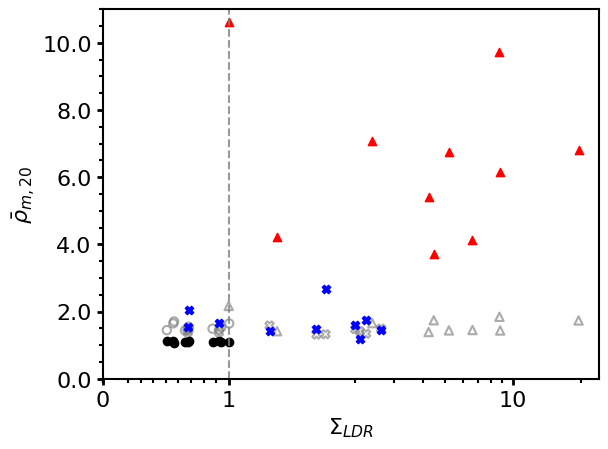}} 

\caption{The mean Mahalanobis density calculated for the \textit{3D} and \textit{6D} phase spaces plotted against other methods of quantifying substructure for 10 subvirial and supervirial simulations which have little to no initial substructured with fractal dimension $D_{\rm f}=3.0$ and 1 pc radii. The left-hand panels show the subvirial results and the right-hand panels show the supervirial regions. The initial values at 0 Myr are represented by the black circles, the blue pluses show 1 Myr and the red triangles show 5 Myr. We show the mean \textit{6D} Mahalanobis densities at 0 Myr, 1 Myr and 5 Myr with grey open circles, grey open crosses and grey open triangles, respectively. From top to bottom the rows show the different methods which $\bar{\rho}_{\rm m, 20}$ is plotted against, with the top row showing $\Lambda_{\rm MSR}$, second row showing $Q$ and the bottom row showing $\Sigma_{\rm LDR}$.}
\label{fig:maha comp 3D1 1pc D=3.0}
\end{figure*}

We now show the same plots but for simulations with little to no primordial spatial or kinematic substructure. Figure~\ref{fig:maha comp 3D1 1pc D=3.0} shows the mean \textit{3D} and \textit{6D} Mahalanobis densities plotted against the established methods for simulations that have an initial fractal dimension $D_{\rm f}=3.0$ and radius 1 pc. The Mahalanobis densities for these simulations increase over time, with supervirial simulations having higher Mahalanobis densities compared to the subvirial simulations after 10 Myr of evolution. 

Panels \textit{(a)} and \textit{(b)} show $\bar{\rho}_{\rm m, 20}$ plotted against $\Lambda_{\rm MSR}$ for the 10 simulations. For the subvirial simulations we see mass segregation detected in three of the 10 simulations, for the supervirial simulations we detect mass segregation in two of the 10 simulations (recall that our threshold for declaring mass segregation is $\Lambda_{\rm MSR} > 2$.)

As for the highly substructured simulations ($D_{\rm f} = 1.6$), the plot of the mean Mahalanobis density when combined with the $Q$-parameter gives the clearest distinction between the different snapshots. For panels \textit{(e)} and \textit{(f)} we show $\bar{\rho}_{\rm m, 20}$ against $\Sigma_{\rm LDR}$. We can see that the 10 most massive stars can end up in a wide range of local surface density ratios The subvirial simulations have a wider range of values, with $\Sigma_{\rm LDR}$ between 0.1 and 10, whereas the supervirial simulations all finish with $\Sigma_{\rm LDR} > 1$.

The grey markers in Figure~\ref{fig:maha comp 3D1 1pc D=3.0} show the \textit{6D} Mahalanobis densities against the established methods. We find once again that the spread in the Mahalanobis densities has decreased, making differentiating between different times or virial states impractical.

\section{Discussion}
\label{sec:discussion}

We have been motivated to test the Mahalanobis density due to its recent applications in quantifying the phase space of exoplanet host stars. In \citet[][]{winter_stellar_2020} they propose that hot Jupiters are more likely to be found around host stars that are in high \textit{6D} phase space density, as measured using the \textit{6D} Mahalanobis density. However, this was questioned by \citet[][]{mustill_hot_2022} who show the peculiar velocities introduce a bias that once accounted for results in no significant excess of hot Jupiters around host stars in high \textit{6D} phase space density. 
The aim of this work is not to make a scientific assessment of the Mahalanobis density in its use in planet formation specifically but simply to see how it changes over time when looking at simple N-body simulations of star-forming regions to see what information, if any, it may be able to give us about the initial conditions (i.e. virial state, density and initial morphology).

Due to the simplicity of our simulations there are a number of important caveats that must be taken into account. First, there is no galactic potential or tidal force acting on our simulated star-forming regions. The presence of an external Galactic tidal field would likely increase the dissolution of the star-forming region by causing outlying stars to become unbound, which would in turn increase the potency of the Galactic tidal field at later ages.

Two more important caveats are that we do not simulate any gas, and therefore there is no gas potential and also that our systems are fully isolated. The most important caveat that disallows direct comparison to the works of \citet{winter_stellar_2020} and \citet[][]{mustill_hot_2022} is that we do not simulate planets in our simulations and so how representative our simulations are of real regions with exoplanet host stars is uncertain. 

In \S~\ref{subsec:comparing to established methods} we show that the \textit{6D} Mahalanobis density in isolation cannot be used to reliably infer the initial conditions of star-forming regions due to overlap in the sub- and supervirial values. However, when the Mahalanobis density in the \textit{3D} phase space is combined with either $\Lambda_{\rm MSR}$, $\Sigma_{\rm LDR}$ or $Q$ then the initial virial conditions can be inferred. This is most clear to see in Figure~\ref{fig:maha comp 3D1 1pc D=3.0}.

The regions that are initially supervirial attain higher final phase space densities than subvirial regions. 

This is somewhat counter intuitive, but is due to the fact that as the region expands small groupings of stars can form which will have similar positions and therefore higher phase space densities. The Mahalanobis distances between these groupings is reduced when we multiply by the inverse of the covariance matrix. In the subvirial cases we see lower \textit{3D} phase space densities due to stars being ejected and ending up in relative isolation compared to the rest of the region. As we are using the mean Mahalanobis density we are sensitive to only a few stars being ejected. 

We cannot determine the initial conditions using the \textit{6D} Mahalanobis density. One would assume that the more data we have (and therefore dimensions in the phase space), the more clearly we would see the distinction between sub- and supervirial simulations. However, somewhat counter intuitively, adding more dimensions to the phase space effectively \textit{'washes'} out any information that would allow us to determine the initial conditions of our star-forming regions. For example, in our $D_{\rm f}=1.6$ supervirial simulations, as the stars dynamically evolve they may get further apart spatially but kinematically they may be quite similar. If two stars that are very far apart end up moving in the same general direction the velocity and positional phases spaces will effectively cancel each other out and therefore removing any information about the initial conditions of the region (i.e. different positions but similar velocities).

We find that the \textit{6D} Mahalanobis density for all simulations is similar at 10 Myr for both sub- and supervirial regions independent of the fractal dimension $D_{\rm f}$ and the initial radii of the region. We therefore suggest that the Mahalanobis distance, and its associated density, are not suitable for quantifying the initial conditions of star formation, nor any subsequent dynamical evolution.

\section{Conclusion}
\label{sec:conclusion}
We present N-body simulations with different initial fractal dimensions, virial states and initial radii and quantify the \textit{3D} and \textit{6D} phase space density using the Mahalanobis distance. We compare the performance of the Mahalanobis density to more established methods for quantifying structure in star-forming regions, namely $\Lambda_{\rm MSR}$, $\Sigma_{\rm LDR}$ and $Q$. We also applied the Mahalanobis distance in \textit{3D} phase space to sets of static synthetic regions of different morphologies to test its ability to discriminate between different morphologies. Our conclusions are as follows:

\begin{enumerate}

    \item The Mahalanobis distance in the \textit{3D} phase space is degenerate across a wide range of morphologies commonly observed in star-forming regions, associations and clusters, and so it cannot be used to differentiate between different morphologies. 
    
    \item The \textit{3D} Mahalanobis densities, $\rho_{\rm m,20}$ can be used to distinguish between the high and low stellar density regions with large amounts of substructure ($D_{\rm f} = 1.6$). The low stellar density regions regions show similar behaviour but delayed by around 0.6 Myr compared to the high volume density regions due to the dynamical timescales being longer. This effect is even more pronounced in the simulations with little to no initial substructure ($D_{\rm f} = 3.0$) where we see the \textit{`bump'} occurring several Myr later than in the higher stellar density simulations corresponding to the subvirial collapse.
    
    \item We show that the Mahalanobis density calculated in the \textit{3D} phase space can be used with the $Q$-parameter, $\Lambda_{\rm MSR}$ or $\Sigma_{\rm LDR}$ to infer information about a region's initial virial state.
    
    \item When using the \textit{6D} Mahalanobis densities we see no significant differences between any of the simulations. Adding more parameters (adding more dimensions) to the phase space suppresses any changes in the Mahalanobis density over time. 
    
\end{enumerate}

We therefore advise against using the Mahalanobis distance as a method to quantify the morphology of star-forming regions due to its degeneracy across both substructured regions and also smooth, centrally concentrated regions. 

When applied to spatial and kinematic phase space (6D), all of its discriminatory power is washed out (similar to the issues encountered when applying the $Q$-parameter to kinematic data, \citet[][]{cartwright_measuring_2009}), and we advocate using combinations of spatial and kinematic metrics instead.

\section*{Acknowledgements}
\label{sec:achnowledgements}
Plots have been generated using Matplotlib 3.3.4 \citep{Hunter:2007}. Numerical results calculated using Numpy 1.20.1 and SciPy 1.9.0 \citep[][]{harris2020array, scipy}. RJP acknowledges support from the Royal Society in the form of a Dorothy Hodgkin fellowship. We wish to thank the anonymous reviewer for their comments which have improved the paper.

\section*{Data Availability}
Data is available on reasonable request to the corresponding author.
 



\bibliographystyle{mnras}
\bibliography{paper_references_final.bib} 




\appendix

\section{Mahalanobis Distance versus Mahalanobis Density}
\label{app: maha distance versus maha density}

\begin{figure*}
\hspace{0.8pt}
\subfigure[$\bar{\rho}_{\rm m, 20}$ vs $\bar{M}_{\rm d}$, subvirial, 1pc]{\includegraphics[width=0.49\linewidth]{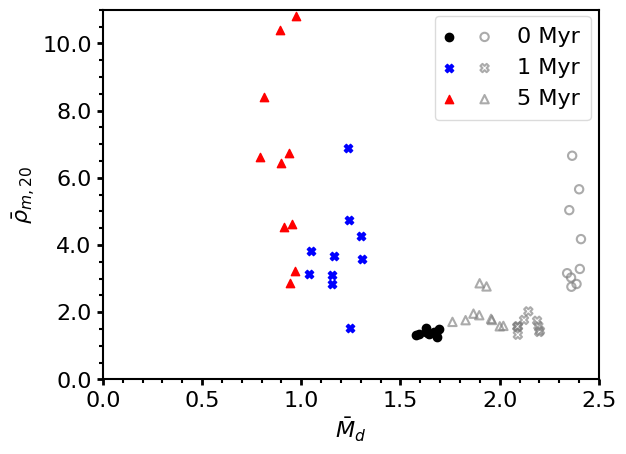}} 
\hspace{0.8pt} 
\subfigure[$\bar{\rho}_{\rm m, 20}$ vs $\bar{M}_{\rm d}$, supervirial, 1 pc]{\includegraphics[width=0.49\linewidth]{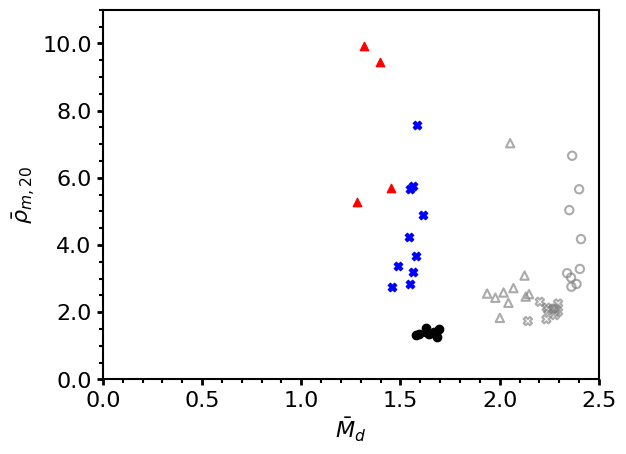}} 

\caption{The mean Mahalanobis density ($\bar{\rho}_{\rm m, 20}$) plotted against the mean Mahalanobis distance ($\bar{M}_{\rm d}$) for highly substructured regions with fractal dimensions $D_{\rm f}=1.6$ and initial radii of 1 pc. Each region contains 1000 stars. The black circles the values at 0 Myr, the blue plus signs are the values at 1 Myr and the red triangles are the values at 5 Myr. The grey open circles, crosses and triangles show the same information but for the Mahalanobis distance and density calculated in the \textit{6D} phase space.}
\label{fig:mahadis vs mahaden all dimensions D=1.6 1 pc}
\end{figure*}

In Figure~\ref{fig:mahadis vs mahaden all dimensions D=1.6 1 pc} we show the relation between the Mahalanobis distance and density across the two different phase spaces investigated and the two different initial virial states.

We see that in the positional phase space (\textit{3D}, the coloured markers) there is significant overlap in both the Mahalanobis distance and density, making differentiating between different snapshots impractical. For the supervirial regions we see less overlap in the Mahalanobis density between the snapshots. However, there is significant overlap between the sub- and supervirial simulations, meaning that neither the Mahalanobis distance nor density can reliably distinguish between different virial states.

\begin{figure*}
\hspace{0.8pt}
\subfigure[$\bar{\rho}_{\rm m, 20}$ vs $\bar{M}_{\rm d}$, subvirial, 1 pc]{\includegraphics[width=0.49\linewidth]{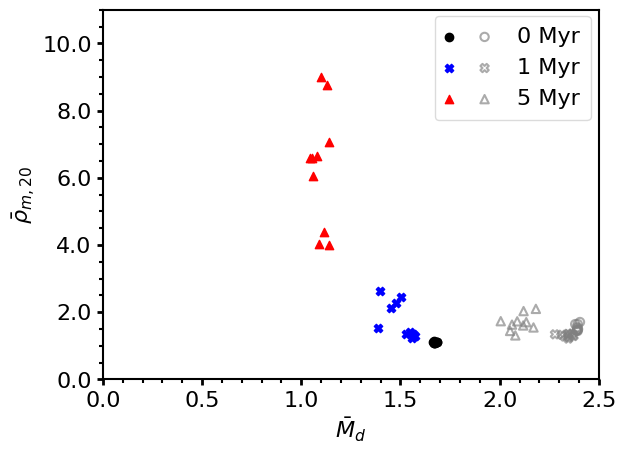}} 
\hspace{0.8pt} 
\subfigure[$\bar{\rho}_{\rm m, 20}$ vs $\bar{M}_{\rm d}$, supervirial, 1 pc]{\includegraphics[width=0.49\linewidth]{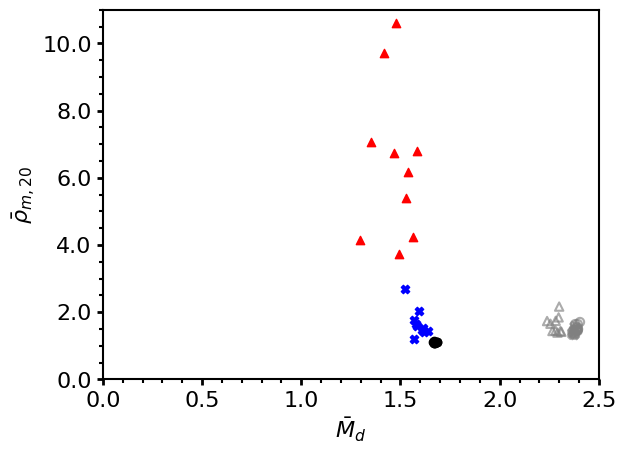}} 

\caption{The mean Mahalanobis density ($\bar{\rho}_{\rm m, 20}$) plotted against the mean Mahalanobis distance ($\bar{M}_{\rm d}$) for substructured regions with fractal dimensions $D_{\rm f}=3.0$ and scales 1 pc for different snapshots. Each region contains 1000 stars. The black circles the values at 0 Myr, the blue crosses are the values at 1 Myr and the red triangles the values at 5 Myr. The grey open circles, crosses and triangles show the same information but for the Mahalanobis distance and density calculated in the \textit{6D} phase space.}
\label{fig:mahadis vs mahaden comp all dimensions 1pc D=3.0}
\end{figure*}

In both Figure~\ref{fig:mahadis vs mahaden all dimensions D=1.6 1 pc} and Figure~\ref{fig:mahadis vs mahaden comp all dimensions 1pc D=3.0} we show the position-velocity phase space (\textit{6D}) with the grey open markers.

Figure~\ref{fig:mahadis vs mahaden comp all dimensions 1pc D=3.0} shows the mean Mahalanobis distance plotted against the mean Mahalanobis density for high density (radii of 1 pc) region with little or no substructure ($D_{\rm f} = 3.0$).

\bsp	
\label{lastpage}
\end{document}